# Detecting Free-Mass Common-Mode Motion Induced by Incident Gravitational Waves: Testing General Relativity and Source Direction via Fox-Smith and Michelson Interferometers

Michael Edmund Tobar[1], Toshikazu Suzuki[2], Kazuaki Kuroda[1]

[1]*Institute of Cosmic Ray Research, University of Tokyo, Tanashi, Tokyo 188, Japan*

[2]*KEK High Energy Accelerator Research Organization, Cryogenics Science Center, 1-1 Oho, Tsukuba-shi, Ibaraki 305, Japan*

In this paper we show that information on both the differential and common mode free-mass response to a gravitational wave can provide important information on discriminating the direction of the gravitational wave source and between different theories of gravitation. The conventional Michelson interferometer scheme only measures the differential free-mass response. By changing the orientation of the beam splitter, it is possible to configure the detector so it is sensitive to the common-mode of the free-mass motion. The proposed interferometer is an adaptation of the Fox-Smith interferometer. A major limitation to the new scheme is its enhanced sensitivity to laser frequency fluctuations over the conventional, and we propose a method of cancelling these fluctuations. The configuration could be used in parallel to the conventional differential detection scheme with a significant sensitivity and bandwidth.

## I. INTRODUCTION

Interferometric gravitational wave detectors are now poised to detect gravitation waves from astrophysical sources over a large detection bandwidth. Large detectors of a few kilometres are currently under construction in Europe and the USA (VIRGO and LIGO) [1-3]. Other current projects include detectors of order a few hundred meters (GEO, TAMA and ACIGA)[4] as well as cryogenically cooled detectors (LCGT)[5]. Standard configurations of these detectors will be sensitive to the quadrupole component of radiation predicted by Einstein's theory of general relativity (GR). Much work has been done in regards to these types of detectors, and for a good description see [6] and references therein.

It is widely acknowledged that GR may not necessarily describe gravity in the strong-field regime, and alternative scalar-tensor theories exist that can not be disproved from experimental evidence to date. Also, it has been shown that these theories may be important in describing inflation models of the universe[7] as well as unified theories such as string theory[8-10]. In particular, it has been shown that Brans-Dicke theory[11] will produce significant amounts of scalar radiation in a collapsing astrophysical systems[12-16], especially in spherical symmetric collapse.

It is well known that spherical resonant-mass detectors can determine the direction of the incoming signal, this principle was first shown in 1971 by Forward[17] and later revived by Merkowitz and Johnson in 1993[18]. Also, it has been known since 1971 that a spherical antenna could be used to distinguish between different polarizations and metric theories of gravitation[17]. Recently a more detailed analysis was performed which has fully revived the concept of the spherical detector [19]. A disadvantage of a spherical detector is that the frequency of detection of any scalar component is limited to the monopole modes of the sphere which are different to the frequency of the spherical quadrupole modes. Thus to determine the quadrupole and scalar content, the radiation itself must be sufficiently broadband to cover both frequencies, which necessarily limits the detection to burst wave forms. Also, a different set of resonant transducers is required at the monopole frequency, and would add to the complexity of the detector.

In this paper the antenna beam patterns were calculated for both the common-mode and differential motion of the interferometer test-masses. Beam patterns were calculated for the six possible polarizations available in metric theories of gravitation. We show from the calculated beam patterns, important information is acquired that can discriminate the direction of the source and between the metric theories of gravitation. Specifically we highlight the example of discriminating between Brans-Dicke scalar waves and Einstein quadrupole radiation. Following this we present a practical scheme based on a Fox-Smith interferometer that may be configured to measure the common-mode response of gravitational radiation at a similar sensitivity and bandwidth to the conventional differential interferometer schemes. The Fox-Smith configuration is not limited to detecting burst sources and is generally broad-band. Also, it could be added as one of the beams in the LIGO detector so a simultaneous detection of the differential and common-mode response to gravitational radiation can be achieved.

## II. INTERFEROMETER RESPONSE TO INCIDENT GRAVITATIONAL WAVES

Incident gravitational radiation will cause relative motion of the two mirror test-masses with respect to the beam splitter of the interferometer. In general this motion will have a differential and common-mode component. Michelson type interferometers are only sensitive to the differential component so past analysis of interferometer response has mainly dealt with the quadrupole radiation of General Relativity causing differential motion of the two test-masses[20][21]. In this section we assume both the common-mode and differential responses may be detected and we calculate the antenna patterns with respect to the six independent possible polarizations [22].

Gravitational waves are believed to propagate as a tensor wave given by;

$$\frac{\partial^2 h_{\alpha\beta}}{\partial x_{\alpha}^2} - \frac{1}{c^2}\frac{\partial^2 h_{\alpha\beta}}{\partial t^2} = 0 \tag{1}$$

The general form from of a gravitational wave in the z-direction can be written as;

$$h_{\alpha\beta} = \begin{pmatrix} h_{Re[\Psi_4]}e^{i\phi_{Re[\Psi_4]}}t_{\alpha\beta} + h_{Im[\Psi_4]}e^{i\phi_{Im[\Psi_4]}}s_{\alpha\beta} + h_{Re[\Psi_3]}e^{i\phi_{Re[\Psi_3]}}q_{\alpha\beta} \\ +h_{Im[\Psi_3]}e^{i\phi_{Im[\Psi_3]}}p_{\alpha\beta} + h_{\Phi_{22}}e^{i\phi_{\Phi_{22}}}n_{\alpha\beta} + h_{\Psi_2}e^{i\phi_{\Psi_2}}m_{\alpha\beta} \end{pmatrix} e^{-i(\omega t - kz)} \tag{2}$$

Here the subscripts follow the Newman-Penrose parameters; $Re[\Psi_4]$ is the plus (or in phase) quadrupole polarization; $Im[\Psi_4]$ is the cross (or quadrature) quadrupole polarization; $Re[\Psi_3]$ is the in phase vector polarization; $Im[\Psi_3]$ is the quadrature vector polarization; $\Phi_{22}$ is the transverse scalar polarization; $\Psi_2$ is the longitudinal scalar polarization. Each polarization in equation (2) is represented by a scalar amplitude, $h_i$, and phase shift, $\phi_i$, followed by a second order tensor that describes the pattern of the polarization. The pattern tensors have the form;

$$t_{\alpha\beta} = \begin{bmatrix} 0 & 0 & 0 & 0 \\ 0 & 1 & 0 & 0 \\ 0 & 0 & -1 & 0 \\ 0 & 0 & 0 & 0 \end{bmatrix} \quad s_{\alpha\beta} = \begin{bmatrix} 0 & 0 & 0 & 0 \\ 0 & 0 & 1 & 0 \\ 0 & 1 & 0 & 0 \\ 0 & 0 & 0 & 0 \end{bmatrix} \quad q_{\alpha\beta} = \begin{bmatrix} 0 & 0 & 0 & 0 \\ 0 & 0 & 0 & 1 \\ 0 & 0 & 0 & 0 \\ 0 & 1 & 0 & 0 \end{bmatrix}$$

$$p_{\alpha\beta} = \begin{bmatrix} 0 & 0 & 0 & 0 \\ 0 & 0 & 0 & 0 \\ 0 & 0 & 0 & 1 \\ 0 & 0 & 1 & 0 \end{bmatrix} \quad n_{\alpha\beta} = \begin{bmatrix} 0 & 0 & 0 & 0 \\ 0 & 1 & 0 & 0 \\ 0 & 0 & 1 & 0 \\ 0 & 0 & 0 & 0 \end{bmatrix} \quad m_{\alpha\beta} = \begin{bmatrix} 0 & 0 & 0 & 0 \\ 0 & 0 & 0 & 0 \\ 0 & 0 & 0 & 0 \\ 0 & 0 & 0 & 1 \end{bmatrix} \tag{3}$$

To calculate the output response we follow closely the method used by Forward where a tensor format for the combined response of the two interferometer arms was assumed[20]. Forward showed the response could be written as;

$$\Delta\xi = \frac{1}{2}h_{\alpha\beta}A^{\alpha\beta} \tag{4}$$

Where $A^{\alpha\beta}$ is the tensor format of the response of the two arms. Assuming that the arms of the interferometer are along the x and y axis and $l$ is equal to the arm length, the differential format can be written as;

$$A^{\alpha\beta}_{diff} = \begin{bmatrix} 0 & 0 & 0 & 0 \\ 0 & 1 & 0 & 0 \\ 0 & 0 & -1 & 0 \\ 0 & 0 & 0 & 0 \end{bmatrix} l \qquad (5)$$

and the common-mode format may be written as;

$$A^{\alpha\beta}_{cm} = \begin{bmatrix} 0 & 0 & 0 & 0 \\ 0 & 1 & 0 & 0 \\ 0 & 0 & 1 & 0 \\ 0 & 0 & 0 & 0 \end{bmatrix} l \qquad (6)$$

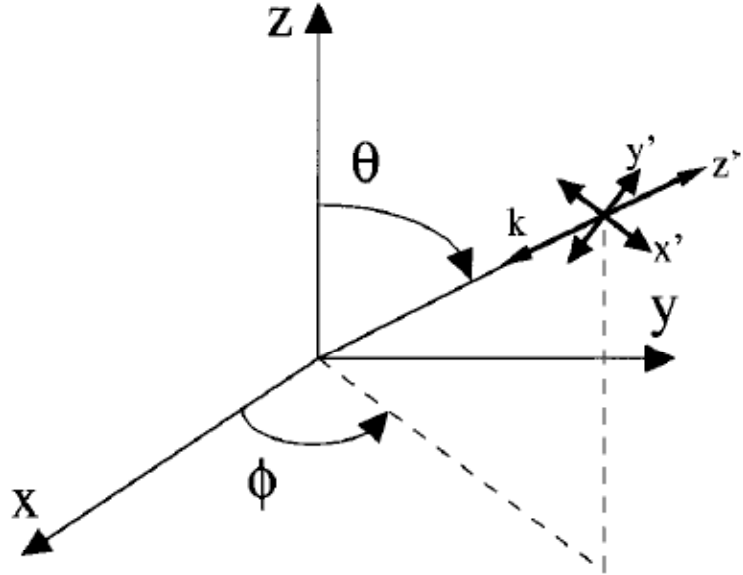

FIG. 1. Co-ordinate system where the x and y axis represent the directions of the interferometer arms, and the dashed system represents the coordinates of the gravitational radiation.

To calculate the sensitivity pattern we assume the gravitational wave is incident on the interferometer from an arbitrary direction defined by angles $\theta$ and $\phi$, but with the polarization angle $\psi$ assumed to be zero, as shown in figure 1. Before the response given by (4) can be calculated $h_{\alpha\beta}$ must be converted to the coordinate system of the interferometer given that $A^{\alpha\beta}$ is in the interferometer frame. To do this we use the general form of the rotation matrix, $R^{\alpha}_{\beta}$, with the Euler angle $\psi$ equal to zero [23];

$$R_\beta^\alpha = \begin{bmatrix} 1 & 0 & 0 & 0 \\ 0 & \cos\theta\cos\phi & \cos\theta\sin\phi & \sin\theta \\ 0 & -\sin\phi & \cos\phi & 0 \\ 0 & -\sin\theta\cos\phi & -\sin\theta\sin\phi & \cos\theta \end{bmatrix} \quad (7)$$

Thus, equation (4) can be rewritten in the interferometer frame as;

$$\Delta\xi_{int} = \frac{1}{2} R_\alpha^{\gamma^{-1}} h_{\gamma\delta} R_\beta^\delta A^{\alpha\beta} \quad (8)$$

We implement this equation by considering the six polarizations of (2) and (3) independently, ie by assuming the amplitude of all the polarizations except for the one under consideration are zero. The normalized (assuming the amplitude is unity) common-mode and differential response per unit arm length of the six polarizations are summarized in table 1, and plotted in figures *2a* to *2l*.

TABLE I. Normalized differential and common mode response of a free-mass interferometer detector per unit arm length to the six possible metric polarizations.

| Radiation type | NP parameter | $\Delta\xi_{int}$ differential | $\Delta\xi_{int}$ common mode |
|---|---|---|---|
| Quadrupole plus (in phase) | $Re[\Psi_4]$ | $\frac{1}{2}\cos 2\phi(1+\cos^2\theta)$ | $\frac{1}{4}(1-\cos 2\theta)$ |
| Quadrupole cross (quadrature) | $Im[\Psi_4]$ | $-\cos\theta\sin 2\phi$ | 0 |
| Vector (in phase) | $Re[\Psi_3]$ | $\sin\theta\sin 2\phi$ | 0 |
| Vector (quadrature) | $Im[\Psi_3]$ | $\frac{1}{2}\cos 2\phi\sin 2\theta$ | $-\frac{1}{2}\sin 2\theta$ |
| Scalar transverse | $\Phi_{22}$ | $\frac{1}{2}\cos 2\phi\sin^2\theta$ | $\frac{1}{2}(1+\cos^2\theta)$ |
| Scalar longitudinal | $\Psi_2$ | $-\frac{1}{2}\cos 2\phi\sin^2\theta$ | $\frac{1}{2}\sin^2\theta$ |

In the past only the differential response to the quadrupole radiation has been considered. In general quadrupole radiation can consist of two polarizations. In our calculations it was assumed that the polarization angle (or Euler angle) was zero. There is no particular angle that is special, as the convention for choosing the zero of the polarization angle is arbitrary. Thus, a polarized gravitational wave will be in fact a linear combination of the Plus and Cross polarizations and the antenna pattern will depend on the angle of polarization. However, if we assume the radiation is unpolarized and consists of many gravitons of random polarization, the antenna pattern may be calculated by taking the square root of the sum of the squares of Re[ψ$_4$] and Im[ψ$_4$]. Fig. 2d shows the response to the unpolarized case, this is the same response calculated by Saulson [6].

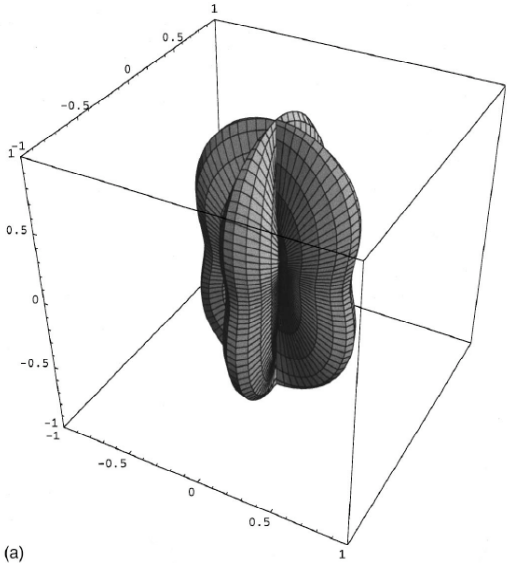

(a)

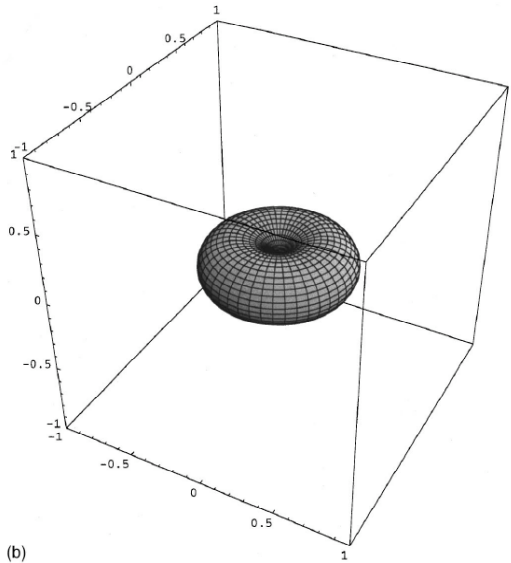

(b)

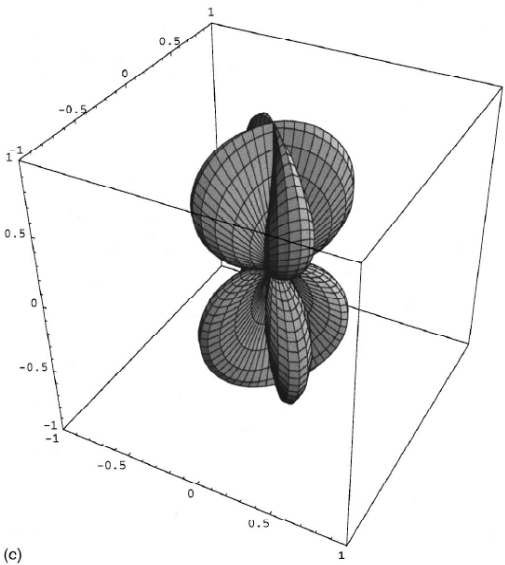

(c)

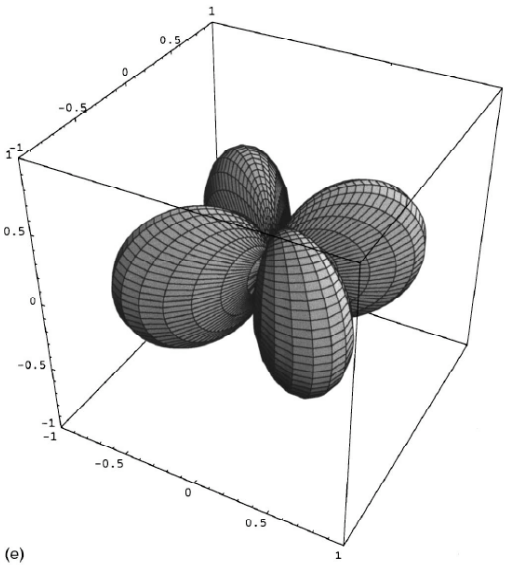

(e)

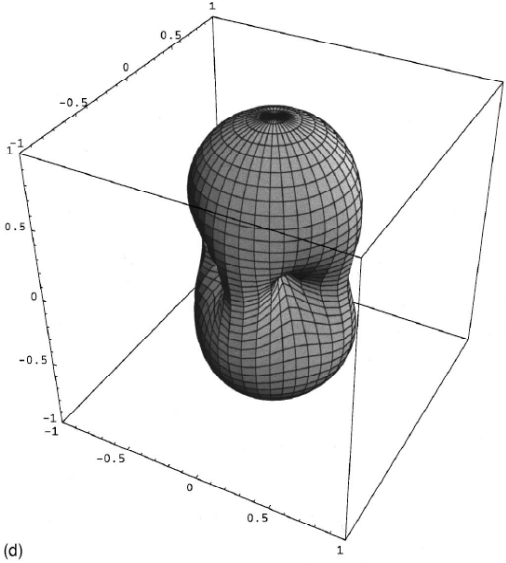

(d)

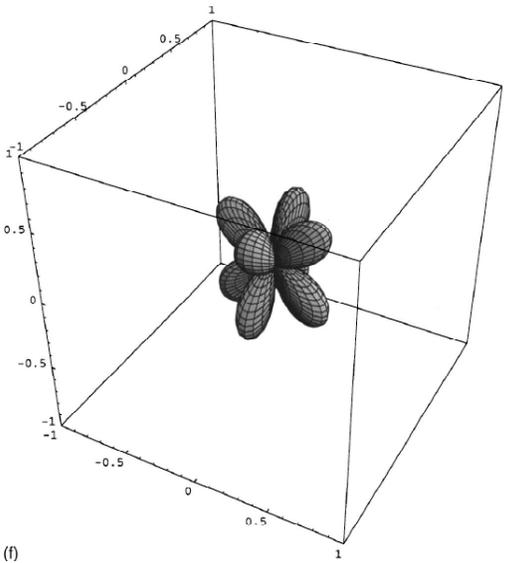

(f)

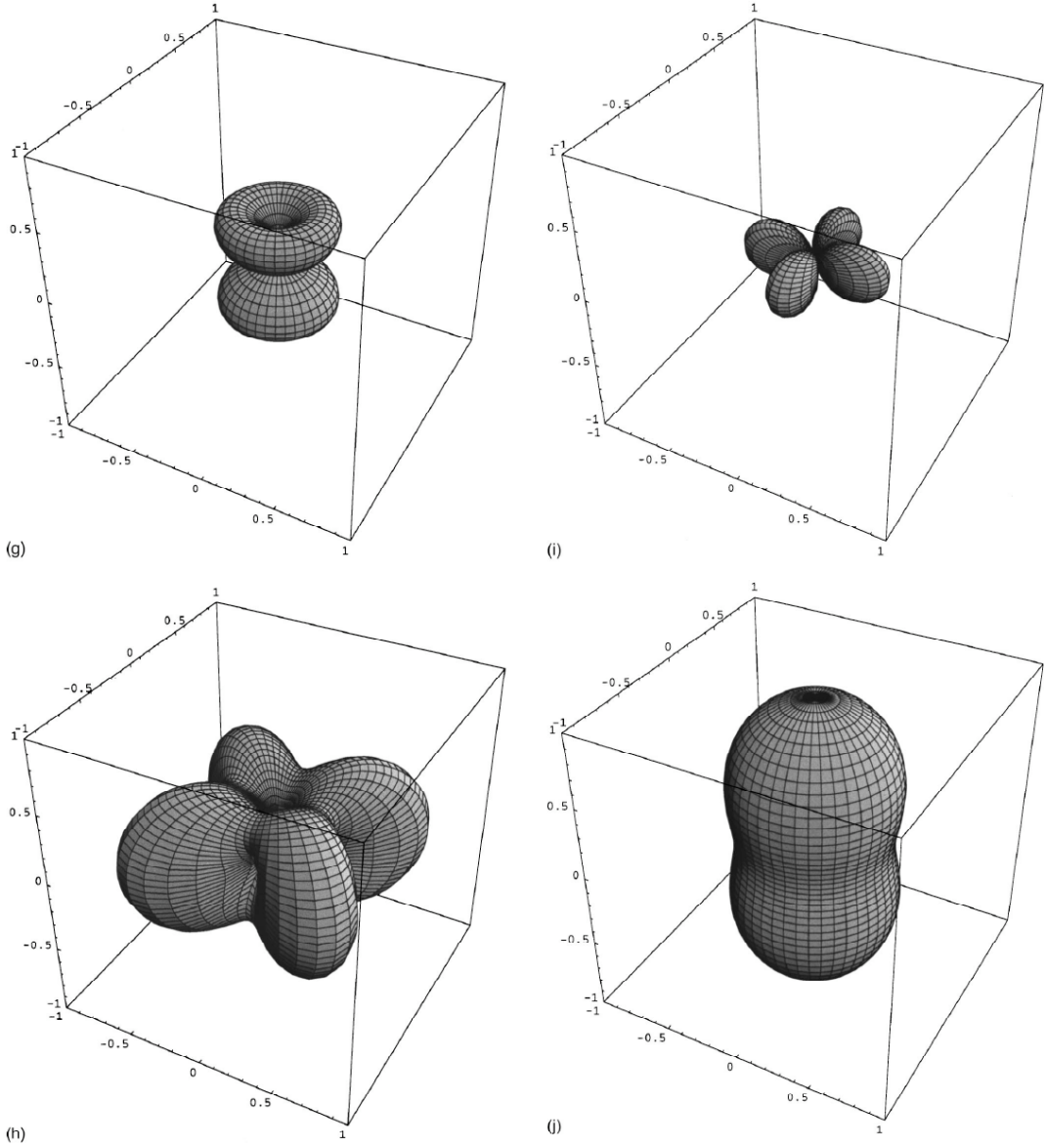

FIG. 2. (Continued).

FIG. 2. (a) Antenna sensitivity pattern for the differential response to the plus quadrupole polarization $\text{Re}[\Psi_4]$, with the $x$-axis labeled on the bottom and the $y$-axis labeled on the top. (b) Antenna sensitivity pattern for the common-mode response to the plus quadrupole polarization $\text{Re}[\Psi_4]$, with the $x$-axis labeled on the bottom and the $y$-axis labeled on the top. (c) Antenna sensitivity pattern for the differential response to the cross quadrupole polarization $\text{Im}[\Psi_4]$, with the $x$-axis labeled on the bottom and the $y$-axis labeled on the top. The common-mode response to this polarization is zero for all $\theta$ and $\phi$. (d) Antenna sensitivity pattern for the differential response to unpolarized quadrupole radiation $(\text{Re}[\Psi_4]^2 + \text{Re}[\Psi_4]^2)^{1/2}$, with the $x$-axis labeled on the bottom and the $y$-axis labeled on the top. The common-mode response to unpolarized radiation is the same as in (b) because the common-mode response to the cross polarization is zero. (e) Antenna sensitivity pattern for the differential response to the in phase vector polarization $\text{Re}[\Psi_3]$, with the $x$-axis labeled on the bottom and the $y$-axis labeled on the top. The common-mode response to this polarization is zero for all $\theta$ and $\phi$. (f) Antenna sensitivity pattern for the differential response to the quadrature vector polarization $\text{Im}[\Psi_3]$, with the $x$-axis labeled on the bottom and the $y$-axis labeled on the top. (g) Antenna sensitivity pattern for the common-mode response to the quadrature vector polarization $\text{Im}[\Psi_3]$, with the $x$-axis labeled on the bottom and the $y$-axis labeled on the top. (h) Antenna sensitivity pattern for the differential response to unpolarized vector radiation $(\text{Re}[\Psi_3]^2 + \text{Re}[\Psi_3]^2)^{1/2}$, with the $x$-axis labeled on the bottom and the $y$-axis labeled on the top. The common-mode response to unpolarized radiation is the same as in (g) because the common-mode response to the in-phase polarization is zero. (i) Antenna sensitivity pattern for the differential response to the transverse scalar polarization $\Phi_{22}$, with the $x$-axis labeled on the bottom and the $y$-axis labeled on the top. (j) Antenna sensitivity pattern for the common-mode response to the transverse scalar polarization $\Phi_{22}$, with the $x$-axis labeled on the bottom and the $y$-axis labeled on the top. (k) Antenna sensitivity pattern for the differential response to the longitudinal scalar polarization $\Psi_{22}$, with the $x$-axis labeled on the bottom and the $y$-axis labeled on the top. (l) Antenna sensitivity pattern for the common-mode response to the longitudinal scalar polarization $\Psi_{22}$, with the $x$-axis labeled on the bottom and the $y$-axis labeled on the top.

The broad angular response of the interferometer to differential motion has been described by Saulson as both a "blessing and a curse". This is because it is very non-directional and behaves more like an ear on the ground than a telescope pointed towards the sky. The blessing is that it is very easy to survey the sky, the curse is that it is very hard to determine the position in the sky without an extreme amount of effort. If only the differential motion is monitored, the direction can be determined from difference in arrival times of signals from detectors at widely separated locations. To uniquely define the position four detectors are needed.

The common-mode response to the plus quadrupole polarization is shown in figure 2b. The response to the cross polarization is zero, thus the response to unpolarized quadrupole radiation will be the same as fig. 2b. The striking feature of the antenna pattern of the common-mode response is that it is very directional and mainly responds to signals in the x-y plane of the interferometer. Thus, if the common-mode response can be detected a comparison with the differential response will give information on the direction of the gravitational wave source.

There is no definite experimental proof that quadrupole radiation is the only type of gravitation radiation. In particular Brans-Dicke theory predicts the existence of a transverse scalar wave. The common-mode response to the transverse scalar wave is shown in fig. 2j. The response is very broad, and thus a sensitive detection scheme for scalar waves can be created by monitoring the common-mode motion. Comparing the differential response of the transverse scalar wave, we note that it is very directional and information regarding the direction of a scalar wave could be determined by monitoring both the common-mode and differential responses.

It is evident that the scalar radiation mainly induces a common-mode signal while the quadrupole radiation mainly induces a differential signal. Thus, by monitoring the relative amounts of each, a test of gravitational theories could be undertaken. There are many combinations of different polarizations that could be looked at. It is not our intention to go through all these possibilities. In the next section we will restrict ourself to Brans-Dicke theory which includes the quadrupole and transverse scalar polarizations.

## III DETERMINING DIRECTION AND THE SCALAR CONTENT IN EINSTEIN AND BRANS-DICKE THEORY

First we assume the theory of General Relativity is correct, and that 100% unpolarized quadrupole radiation is incident on the detector. The ratio of the differential to common-mode response is given by;

$$\gamma_{quad} = \sqrt{\frac{\left(\frac{1}{2}\cos 2\phi \left(1+\cos^2\theta\right)\right)^2 + \left(-\cos\theta \sin 2\phi\right)^2}{\left(\frac{1}{4}(1-\cos 2\theta)\right)^2}} \qquad (9)$$

This function is plotted as a two-dimensional contour plot in fig. 3, with $\theta$ as the vertical axis an $\phi$ as the horizontal. The distinguishing features is that most of the time the differential response is greater than the common-mode response ($\gamma_{quad}>0$ dB), the exception is at the bisectors of the interferometer arms in the x-y plane at $\theta=\pi/2$ and $\phi=\pm\pi/4$ or $\pm 3\pi/4$. At these points the differential response is zero. The common-mode signal remains greater than the differential at all angles in the x-y plane ($\theta=\pi/2$) except along the interferometer arms where the response is equal ($\gamma_{quad}=0$ dB). If the differential response is greater than the common mode response by at least 10 dB this means that $\theta<\pi/4$ or $\theta>3\pi/4$, and as $\theta->0$ or $\theta->\pi$, the common mode response decreases very rapidly to zero. Clearly if we know that the radiation is quadrupole we can determine information about the direction. For example, if $\gamma_{quad}$ was measured to be 10 dB, from figure 3 it could be determine that $\theta$ must be $\pi/4$ or $3\pi/4$. This will describe two circles in the celestial sphere and will reduce the number of necessary detectors to pinpoint the position. If a second detector that uses both common-mode and differential detection was used, the intersection of the four circles on the celestial sphere could be used to pin point at most 8 possible patches on the sky (could be less depending on the orientation of the two detectors with respect to the gravitational wave). If we then used the time difference of arrival between the two detectors, the patch from which the radiation came from could be determined as long as the time difference circle on the celestial sphere lined up with only one of the patches. Thus to determine the direction only two detectors are necessary rather than four. In effect still four detectors are being used ie. two common-mode and two differential. For more details on determining direction and waveform using the conventional interferometer, see Saulson (pg. 255)[6] and Gürsel and Tinto[24].

In Brans-Dicke theory a large amount of transverse scalar radiation can exist as well as the quadrupole component, especially in the case of symmetrical collapse. Suppose now that a pure transverse scalar wave was incident on the detector, the ratio of the differential to common-mode response is then given by;

$$\gamma_{scal} = \sqrt{\frac{\left(\frac{1}{2}\cos 2\phi \sin^2\theta\right)^2}{\left(\frac{1}{2}(1+\cos^2\theta)\right)^2}} \qquad (10)$$

A contour plot of this function is shown in figure 4. For this case the common-mode response is always greater than the differential response, except for when the radiation is incident along the x or y axis of the interferometer, at these points they are equal.

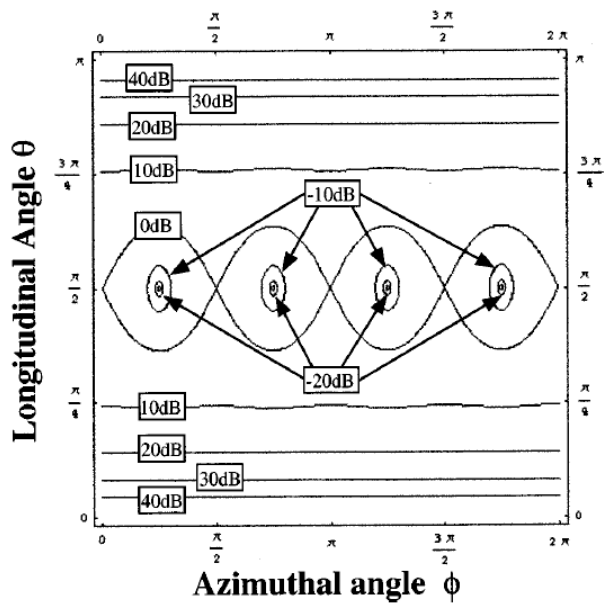
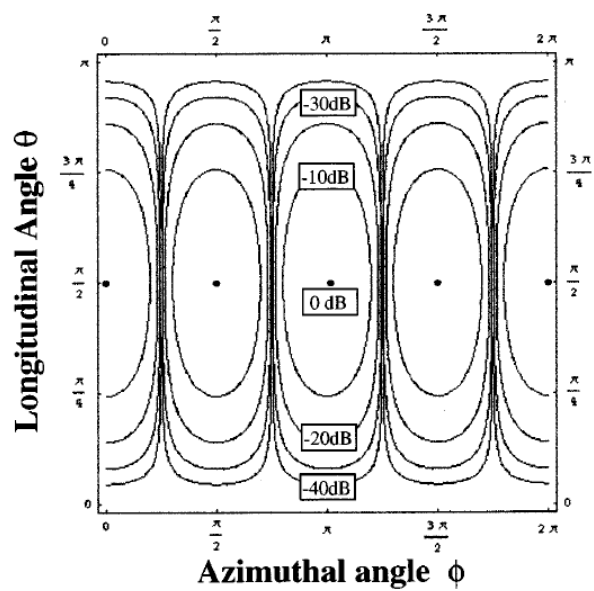

FIG. 3. Contour plot of $\gamma_{quad}$ as a function of azimutal angle $\phi$ and longitudinal angle $\theta$.

FIG. 4. Contour plot of $\gamma_{scal}$ as a function of azimutal angle $\phi$ and longitudinal angle $\theta$.

For a specific value of $\gamma_{scal}$ four possible contours must be considered. In general the contour plot is different to the quadrupole plot of figure 3. The only plane where they are equal is in the x-y plane. Thus, if the direction of the gravitational wave source can be determined using the time difference technique of four detectors, then the amount of differential to common-mode response could uniquely determine the scalar and quadrupole content of the radiation. The exception is in the x-y plane where a transverse unpolarized quadrupole gravitational wave will give the same response as the transverse scalar wave.

It is clear that a major benefit can be made if the common-mode response of a free-mass interferometer is monitored along with the differential response. In the next section we introduce a scheme based on a Fox-Smith interferometer that in principle can measure the common-mode response with a significant sensitivity.

## IV. FOX-SMITH CONFIGURATION

The basic adaptation for the Fox-Smith (F-S) configuration is shown in Fig. 5. It is obtained by simply rotating the beam splitter of the normal configuration by $90^0$. However, the operation is somewhat different. The conventional interferometer is non-resonant unless Fabry-Perot (F-P) cavities are purposefully put in the arm of the detector. In this case it is not necessary to put mirrors in the arms as the beam splitter forms a L-shaped F-S cavity with the two end mirrors.

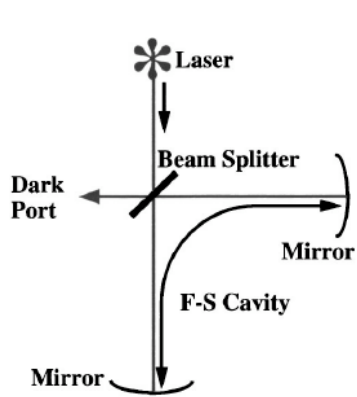

FIG. 5. Schematic of the Fox-Smith (F-S) interferomeric detector.

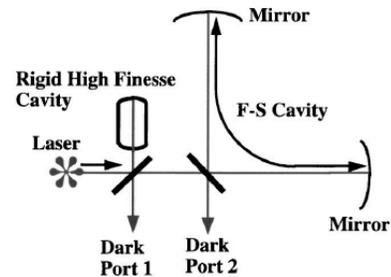

FIG. 6. Fox-Smith Interferometer with phase noise cancelation. Dark Port 1 and 2 are both sensitive to the common-mode motion of the mirrors. However, dark port 2 is highly sensitive to the laser frequency noise, while dark port 1 is not. Dark port 2 gives a highly sensitive way of monitoring the performance of the laser while the system is online. The rigid high finesse cavity could be either Fabry-Pérot (F-P) or (F-S).

Quadrupole gravitational waves will induce mostly differential shifts in the arm lengths of an interferometric gravitational wave antenna, depending on the direction of the incoming wave. For this reason the conventional interferometric detector is configured to measure the differential motion of the arm lengths. However, for the new configuration shown in Fig. 5, the path length of the laser light in the F-S interferometer will be directly perturbed by common mode changes in the arms and remain unchanged with respect to differential changes. A scalar gravitational field will induce mainly common mode shifts in the arms. Hence the system in Fig. 5 will be mainly sensitive to scalar gravitational radiation as discussed previously.

The interferometer is also sensitive to the frequency noise of the incident laser as it interferes light directly reflected from the interferometer with resonant light. In fact this scheme is the laser equivalent of the microwave interferometric read out used on the Niobe resonant-mass detector at UWA, which is also sensitive to the pump oscillator frequency noise [25]. Well known ways of cancelling the phase noise at microwave frequencies may be adapted to laser frequencies. One such method is to implement a tuned ridged dummy cavity that can supply the same dispersion as the resonant motion sensor. This method has been successfully used at microwaves in a high-sensitivity room temperature sapphire transducer [26]. The proposed F-S laser interferometer with phase noise cancellation is shown in Fig. 6. Like the conventional configuration, we will show that the output port can be tuned to be dark with all the power reflected back to the laser. This is imperative for the read out electronics to operate in the non-saturated small-signal regime where the noise added is a minimum. In the following sections mathematical analysis is presented to show the principles of operation and to compare the sensitivity to the conventional Michelson Fabry-Perot schemes.

## V. MATHEMATICAL ANALYSIS

### A. Basic Fox-Smith interferometer

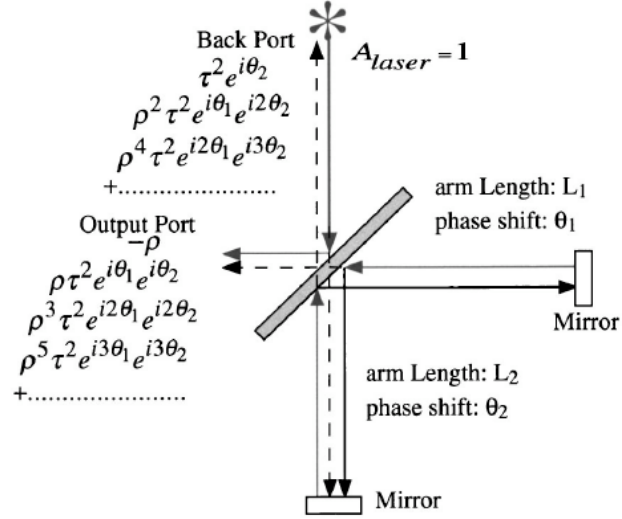

FIG. 7. Ray diagram of the Fox-Smith interferometer.

To illustrate the principle of operation the following assumptions are made; 1. A normalized laser amplitude of zero phase shift is incident on the interferometer; 2. the lengths of the interferometer arms are $L_1$ and $L_2$, with a corresponding phase shift of $\theta_1$ and $\theta_2$; 3. perfect reflectivity mirrors exist at the end of the arms; 4. the beam splitter phase reference is taken to give a reflection coefficient of $-\rho$ and $\rho$ from the incident laser side and F-S cavity side respectively, with a transmission coefficient of $\tau$. From these assumptions a typical zig-zag diagram that traces the laser rays throughout the interferometer is shown in Fig. 7. The amplitude at the output and back port can be summed up to give;

$$A_{out} = \rho\left(-1 + \tau^2 e^{i\theta_2} e^{i\theta_1}\left(1 + \rho^2 e^{i\theta_2} e^{i\theta_1} + \rho^4 e^{i2\theta_2} e^{i2\theta_1} + ......\right)\right) \quad (11a)$$

$$A_{back} = \tau^2 e^{i\theta_2}\left(1 + \rho^2 e^{i\theta_2} e^{i\theta_1} + \rho^4 e^{i2\theta_2} e^{i2\theta_1} + ......\right) \quad (11b)$$

Assuming a lossless system where $\tau^2=1-\rho^2$, and given that $(1-x)^{-1}=1+x+x^2+x^3+...$, then (11) can be rewritten as;

$$A_{out} = \rho\left(\frac{e^{i\theta_1} e^{i\theta_2} - 1}{1 - \rho^2 e^{i\theta_1} e^{i\theta_2}}\right) \quad (12a)$$

$$A_{back} = \frac{(1-\rho^2)e^{i\theta_2}}{1-\rho^2 e^{i\theta_1} e^{i\theta_2}} \quad (12b)$$

For the interferometer to be useful as a detector the rays of the laser light must interfere in a way to give destructive interference and a null output, with $A_{out}$ equal to zero and $A_{back}$ equal to unity. To satisfy this situation the following phase condition must be met,

$$\theta_1 = 2n_1\pi \quad \theta_2 = 2n_2\pi \quad (13)$$

where $n_1$ and $n_2$ are the integral number of wavelengths in the interferometer arms. This is a similar requirement for a conventional interferometer which also must have its phase adjusted to operate on a null. Condition (13) also satisfies the necessary condition that the laser is incident at a resonant frequency.

The dark output is independent on the reflection coefficient, and hence the ratio in which the beam is split is on first glance not important (later we show it does have implications for the sensitivity). Just as the dark port output is independent of beam splitter reflectivity, so is the reverse amplitude $A_{back}$, and is equal to unity. When the reflectivity of the mirror is small the power in arm 2 is approximately equal to the laser power, and the power in arm 1 approaches zero. This limit approximates a single arm delay line interferometer with one pass. As the reflectivity increases the finesse of the F-S cavity increases and so does the power in the interferometer arms. Thus, one might expect the sensitivity of the configuration to increase with beam splitter reflectivity. This is indeed true and later we show how the sensitivity depends on reflectivity.

### B. Signal sensitivity to interferometer phase changes

Gravitational waves interacting with the free-mass system will cause small perturbations in the length and hence phase of the interferometer arms. A differential-mode length change of two arms of similar length will cause the phase given by (13) to be perturbed, and can be written as;

$$\theta_1 = 2n_1\pi + \Delta\theta \quad \theta_2 = 2n_2\pi - \Delta\theta \quad (14)$$

where $\Delta\theta$ is the phase perturbation in each arm. Substituting (14) into (12) gives $A_{out}=0$ and $A_{back}=1$, and shows the F-S scheme is insensitive to differential motion of the end mirrors.

On the other hand, a common-mode shift of the two arms requires the phase shift of the two arms to be given by;

$$\theta_1 = 2n_1\pi + \Delta\theta \quad \theta_2 = 2n_2\pi + \Delta\theta \tag{15}$$

Substituting into (12a) and assuming $\Delta\theta<<1$, then the corresponding change in amplitude at the dark port can be calculated to be;

$$\Delta A_{out} = \frac{2i\rho\Delta\theta}{1-\rho^2} \tag{16}$$

From (16) the resonant effect is apparent as the beam splitter reflectivity increases so does the magnitude of the signal. For this configuration it is impossible to tell between frequency fluctuations of the laser and of the F-S cavity, thus the phase noise of the laser will be amplified by the same amount as the signal, and will limit the sensitivity.

## C. Fox-Smith interferometer with phase noise cancellation

In this section we illustrate how the phase noise can be cancelled by interfering the bright port of the F-S cavity with a rigid F-P cavity, while remaining sensitive to the common mode displacement of the F-S cavity arms. In actual fact the F-P cavity will exhibit resonant frequencies inversely proportional to its length and will not be ideally rigid. This effect is discussed in detail later.

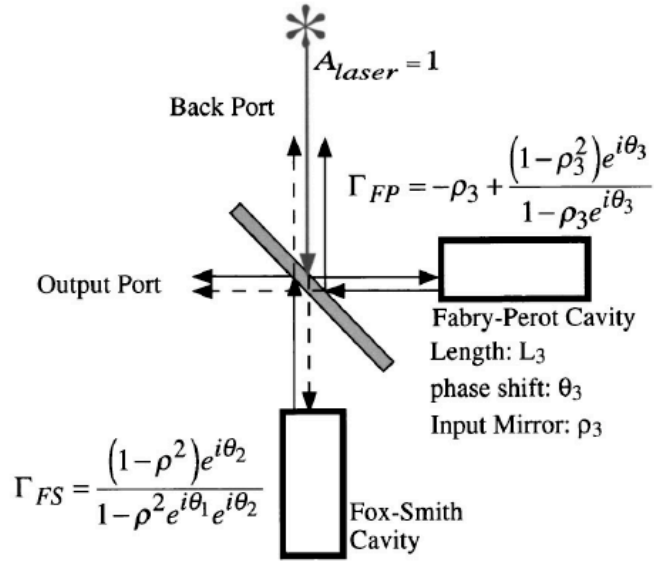

FIG. 8. Interference of light from a Fox-Smith free-mass interferometer with a rigid Fabry-Pérot cavity. The complex reflection coefficient from both cavities, $\Gamma_{FS}$ and $\Gamma_{FP}$ are given assuming the end mirrors are highly reflective compared to the input mirrors.

The schematic of the set up is shown in fig. 8. Assuming a lossless beam splitter with a reflection coefficient $\rho_{bs}$ and $-\rho_{bs}$ on the laser side and F-S cavity side respectively, with a transmission of $\sqrt{1-\rho_{bs}^2}$, the amplitude at the output port is given by;

$$A_{out} = \rho_{bs}\sqrt{1-\rho_{bs}^2}\left(\frac{\left(1-\rho_3^2\right)e^{i\theta_3}}{1-\rho_3 e^{i\theta_3}} - \rho_3 - \frac{\left(1-\rho^2\right)e^{i\theta_2}}{1-\rho^2 e^{i\theta_1}e^{i\theta_2}}\right) \tag{17}$$

To be tuned on the cavity resonant frequencies condition (13) for the F-S cavity must be met, along with the following condition the for the F-P cavity;

$$\theta_3 = 2m\pi \tag{18}$$

where m is an integer. Substituting (18) and (13) into (17) gives $A_{out}=0$ which means it functions as a dark port. For the phase noise to cancel the dispersion of the laser phase reflected from both cavities with respect to frequency must be equal.

### D. Phase response of the F-S and F-P cavities

To calculate the phase response the argument of the reflection coefficients are differentiated with respect to the phase lengths of the cavities and calculated on resonance. For the F-P cavity the argument of $\Gamma_{FP}$, $\phi_{FP}$, is differentiated with respect to $\theta_3$ to give;

$$\left.\frac{d\phi_{FP}}{d\theta_3}\right|_{\theta_3=2m\pi} = \frac{1+\rho_3}{1-\rho_3} \tag{19}$$

Likewise the argument of $\Gamma_{FS}$, $\phi_{FS}$, is differentiated with respect to $\theta_T = \theta_1 + \theta_2$ to give;

$$\left.\frac{d\phi_{FS}}{d\theta_T}\right|_{\theta_T=2(n_1+n_2)\pi} = \frac{1}{2}\left(\frac{1+\rho^2}{1-\rho^2}\right) \tag{20}$$

The phase shift in the arms of the F-P and F-S cavity are related to the length by $\theta_3 = -2kL_3$ and $\theta_T = -2k(L_1+L_2)$, where $k=2\pi f/c$, where c is the speed of light and f is the laser frequency. Thus the dispersion of the reflected light with respect to frequency of the laser when incident at the resonant frequency can be calculated to be;

$$\frac{d\phi_{FP}}{df} = -\left(\frac{1+\rho_3}{1-\rho_3}\right)\left(\frac{4\pi}{c}\right)L_3 \tag{21a}$$

$$\frac{d\phi_{FS}}{df} = -\left(\frac{1+\rho^2}{1-\rho^2}\right)\left(\frac{4\pi}{c}\right)\left(\frac{L_1+L_2}{2}\right) \tag{21b}$$

For the phase noise to cancel the dispersion relations given by (21) must be designed to be equal. This can be achieved by designing the length and reflectivity of the cavities appropriately.

### E. Condition for phase noise cancellation

By equating (21a) and (21b) the relation between the reflection coefficient of the input beam splitter of the F-S cavity, $\rho$, and the input mirror of the F-P cavity $\rho_3$ can be shown to be;

$$\rho_3 = 1 - \frac{2\gamma(1-\rho^2)}{1+\rho^2+\gamma(1-\rho^2)} \tag{22}$$

where $\gamma=2L_3/(L_1+L_2)$ is the length ratio between the average arm length of the F-S cavity and the length of the F-P cavity. To calculate the Finesse of the required F-P cavity the following formula can then be used;

$$F_{FP} = \frac{\pi\sqrt{\rho_3}}{1-\rho_3} \tag{23}$$

Where $F_{FP}$ is the cavity finesse.

### F. Signal Phase Sensitivity

Gravitational waves interacting with the free-mass system will cause small perturbations in the F-S interferometer arms, but will not change the path length significantly of the rigid F-P cavity. Assuming a small differential motion of the two arms of the F-S cavity, the phase given by (13) and (18) may be rewritten as;

$$\theta_1 = 2n_1\pi + \Delta\theta \quad \theta_2 = 2n_2\pi - \Delta\theta \quad \theta_3 = 2m\pi \tag{24}$$

where $\Delta\theta$ is the phase perturbation in each arm. Substituting (24) into (17) gives the change in amplitude at the dark port to be, $\Delta A_{out} \approx 0$, for a high finesse system, and shows the scheme is insensitive to differential motion of the end mirrors.

On the other hand, a common-mode shift of the two arms requires the phase shift of the cavities to be given by;

$$\theta_1 = 2n_1\pi + \Delta\theta \quad \theta_2 = 2n_2\pi + \Delta\theta \quad \theta_3 = 2m\pi \tag{25}$$

Substituting (25) into (17) and assuming $\Delta\theta \ll 1$, then the corresponding change in amplitude at the dark port can be calculated to be;

$$\Delta A_{out} = -i\rho_{bs}\sqrt{1-\rho_{bs}^2}\left(\frac{1+\rho^2}{1-\rho^2}\right)\Delta\theta \tag{26}$$

This has maximum sensitivity when the beam splitter reflection coefficient is $\rho_{bs}=1/\sqrt{2}$. Thus, for a high finesse system where $\rho \to 1$, $|\Delta A_{out}| \approx \Delta\theta/(1-\rho^2)$, which is a factor of two smaller than the F-S system without phase noise cancellation given by (16). This is the price to pay to cancel the phase noise.

# VI. SENSITIVITY COMPARISON TO THE CONVENTIONAL DIFFERENTIAL GRAVITATIONAL WAVE INTERFEROMETER

## A. Single pass Michelson interferometer

First we consider the conventional single pass gravitational interferometer as shown in Fig. 9. Here the laser beam is split in two and interfered at the output after they traverse through the two independent interferometer arms. The phase difference between the two signals must be exactly equal to cancel the phase noise at the output of the interferometer ie. $\theta_1 = \theta_2$ in fig. 9. At this setting it is also evident that the expression for the output in Fig. 9 goes to zero, and is thus dark.

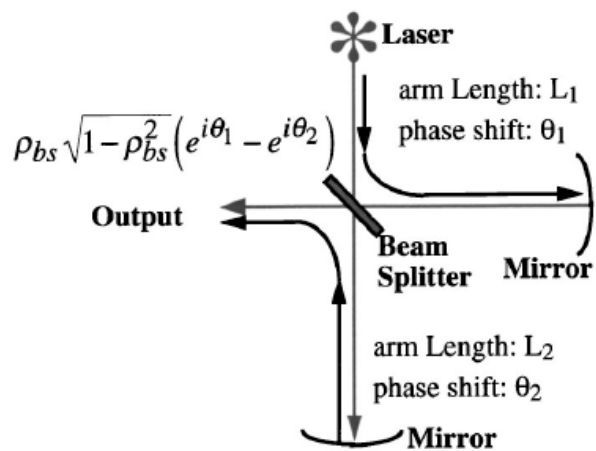

FIG. 9. Single pass Michelson Interferometer.

This detector is sensitive to differential changes in the arm length and insensitive to common mode changes. If we assume that a very small phase shift is introduced, due to a arm of length L changing corresponding to a phase shift of $\pm \Delta \theta$ in arm 1 and arm 2 respectively, then the change in the output amplitude will be given by;

$$\Delta A_{out} = 2i \rho_{bs} \sqrt{1 - \rho_{bs}^2} \, e^{i\theta_1} \Delta \theta \qquad (27)$$

This is maximized when $\rho_{bs}$ is equal to $1/\sqrt{2}$, ie the power split provided by the beam splitter is equal, and in this case $|\Delta A_{out}| = \Delta \theta$.

## B. Fabry-Perot Michelson interferometer

Most proposed large scale interferometers will use F-P cavities in the interferometer arms to enhance the sensitivity. For this configuration the input mirror reflectivity and phase length of both arms must be well matched to cancel the phase noise at the output port. The form of the reflection coefficient from a F-P cavity is shown in Fig. 8. If we add two F-P cavities in the arms of the Michelson interferometer shown in Fig. 9 the amplitude at the output port can be shown to be;

$$A_{out} = \rho_{bs}\sqrt{1-\rho_{bs}^2}\left(\frac{(1-\rho_{fp}^2)e^{i\theta_1}}{1-\rho_{fp}e^{i\theta_1}} - \frac{(1-\rho_{fp}^2)e^{i\theta_2}}{1-\rho_{fp}e^{i\theta_2}}\right) \quad (28)$$

Here it is assumed that the two input mirrors of the F-P cavities have the same reflection coefficient of $\rho_{fp}$. From (28) it is evident that if the phases of the two arms are matched then (28) equals zero and the output port is dark. To calculate the signal sensitivity we assume the phases of the arm lengths are perturbed around the resonant frequencies with a differential signal so that $\theta_1 = 2n\pi + \Delta\theta$ and $\theta_2 = 2n\pi - \Delta\theta$. Substituting into (28) gives;

$$\Delta A_{out} = 2i\rho_{bs}\sqrt{1-\rho_{bs}^2}\left(\frac{1+\rho_{fp}}{1-\rho_{fp}}\right)\Delta\theta \quad (29)$$

Again this is maximised when the beam splitter reflection coefficient is $1/\sqrt{2}$, and is given by; $|\Delta A_{out}| = (1+\rho_{fp})/(1-\rho_{fp})\Delta\theta$. In this case, the sensitivity compared to the single pass Michelson Interferometer has increased by the phase response of the F-P cavity (given by Eqn. 19 ).

### C. Requirements to detect common-mode induced signals at a similar sensitivity as LIGO

The LIGO detector will operate as a Michelson interferometer with matched F-P cavities in each arm. Thus the sensitivity of the signal is governed by (29). To compare the sensitivity we assume that the noise sources for the F-S interferometer are identical and hence we only compare the signal response. This comparison assumes the phase noise cancellation cavity is rigid, the effects of its non-rigidity are discussed in detail later.

The initial LIGO specifications require the input mirrors to have a reflectivity of order 0.97 [27], which corresponds to F-P cavities with a finesse of about 200 and storage time of order 1 ms. From (29) the signal will be amplified by a factor of 65, due to the F-P cavities. To achieve the same amplification factor in the F-S interferometer, from (26), a beam splitter reflectivity of 0.9924 is required. Following this, the reflectivity and finesse of the rigid dummy cavity necessary to cancel the phase noise can be calculated from (22) and (23). For example, given that the LIGO arm length is 4 km, a 8 m cavity will require a cavity finesse of 100 000, which can be easily obtained. Cavity finesse of up to 2 million have been measured previously[28], and the required finesse scales inversely to the length. In this case a length of only 0.4 m would be required. Practically the length could be adjusted to give the required phase noise cancellation. Also, in such a high finesse cavity not all of the power may be reflected. To compensate this the transmitted and reflected power could be recombined before interference in the beam splitter. Like the conventional scheme, the F-S interferometer can be fitted with a power recycling mirror at the laser input port, and a signal recycling mirror at the output port.

### D. Implications of Longitudinal Resonances in the Rigid F-P Cavity

The frequency noise cancellation F-P cavity would have similarities to a resonant bar detector. First it must be vibration isolated and second it will have longitudinal resonances that depend on its length. However, for the structure to work as we intend it must seem rigid and therefore the first longitudinal resonances must be higher than the frequencies of detection. Thus we have two contradictory requirement; 1. the length of the rigid cavity must be long to create the same phase shift as the F-S interferometer; 2. the rigid cavity must be kept short to keep the first longitudinal resonance above the detection frequency of interest. In this section we analyse the relation between the required F-P cavity finesse with respect to the length and resonant frequency of the cavity.

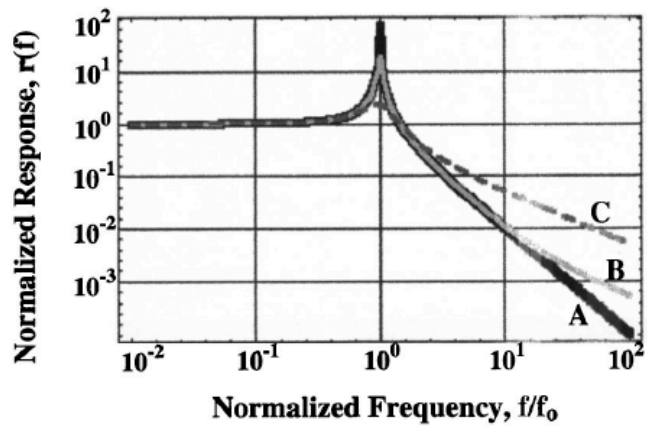

FIG. 10. Normalized output of the response of an incident gravitational wave causing a common mode excitation in the F-S interferometer with a rigid F-P cavity for phase noise cancelation as shown in Fig. 8. A—when the fundamental longitudinal resonance has a $Q \to \infty$; B—when the $Q$ is reduced to 10; C—when the $Q$ is reduced to 1.

Our analysis takes into consideration the lowest resonant frequency of our rigid structure, namely the first longitudinal resonance. It is usual to model the first longitudinal resonance by two masses of half the total mass, joined together by a spring. It is well known that near the resonant frequency of a bar detector there is enhanced sensitivity to gravitational waves. Combining the mass spring model with the technique to calculate the response introduced in section II, it is easy to show that the response of the bar to unpolarized Einstein quadrupole and Brans Dicke scalar radiation is the same. Typically the antenna pattern is a doughnut shape with the maximum sensitivity for directions orthogonal to the longitudinal axis of the bar, and zero sensitivity along the longitudinal axis of the bar. The spectral displacement sensitivity between the two ends of the bar with respect to frequency, $\omega$, may be written as[29];

$$X_1(\omega) = \frac{l_1 \omega^2 H(\omega)}{\left(\omega^2 - \omega_o^2\right) - j\frac{\omega}{\tau}} \tag{30}$$

Here $H(\omega)$ is the strain signal density, $l_1$ is the effective length, $\omega_0$ is the resonant frequency and $\tau$ the decay time of the resonance. For an interferometer of arm length L the displacement as a function of frequency is equal to;

$$X_2(\omega) = l_2 H(\omega) \tag{31}$$

The above is only true for one pass of laser light, given that we must make the optical path length $l_{opt}$ equal in both cases then the normalized response, $\tilde{r}(\omega)$, measured at the output of the interferometer can be calculated from (30) and (31) to be;

$$\tilde{r}(\omega) = \frac{X_2(\omega) - X_1(\omega)}{H(\omega) l_{opt}} = \frac{\omega_o^2 + j\frac{\omega}{\tau}}{\omega_o^2 - \omega^2 + j\frac{\omega}{\tau}} \tag{32}$$

A plot of (32) is shown in figure 10.

From figure 10 it is evident that below the resonant frequency the response from (30) is small and the main component of (32) comes from (31). For frequencies in this regime the interferometer is sensitive to the common mode motion of the free masses of the F-S interferometer. At the resonant frequency the response is enhanced due to the resonant nature of (30). This maybe useful to detect gravitational waves, however, the response can not distinguish whether or not the incident radiation is quadrupole or scalar. Therefore, to utilize the interferometer as suggested in this paper, it may be prudent to damp the motion of the resonance to artificially reduce the Q factor and the size of the resonant motion, and thus its effect. Above resonance the mass behaves like a free mass (ignoring other resonances of course). This is not useful as at these frequencies the gravitational wave signal will experience cancellation as well as the laser phase noise.

The material must be chosen to have low losses to keep the internal thermal noise low, and have a high sound velocity, v, to keep the longitudinal resonance high. The resonant frequency of the bar mainly depends on the length and may be approximated by;

$$\omega_o = \frac{\pi v}{l_1} \qquad (32)$$

Therefore if we chose a material such as sapphire with v=10 km/s then for a 0.4 m long cavity the longitudinal resonant frequency will be 12.5 kHz. It is easy to obtain boules of sapphire this long. Thermal noise in the vicinity of the resonance will be a problem, however sapphire has an extremely high-Q and has been shown to be an ideal test mass for an interferometer[30, 31]. Also, its properties are further enhanced on cooling, ie. its Q increases[32] and so does its dimensional stability[33]. For this reason it has been proposed to cool sapphire test masses in an interferometer[5]. The rigid cavity may be housed in a vacuum chamber independent of the interferometer, and could be cooled and vibrationally isolated in a similar way to a resonant bar detector. This would ensure low noise operation. An elegant way to cold damp the acoustic resonance in a sapphire bar without adding noise is to excite microwave whispering gallery modes in the circumference of the sapphire and use the parametric interaction between the resonances[34]. A plot of the cavity finesse as a function of the fundamental longitudinal resonance for niobium, aluminium 5056 and sapphire is shown in figure 10. These materials have all been found to have good acoustic properties and can be considered to make the noise cancellation cavity.

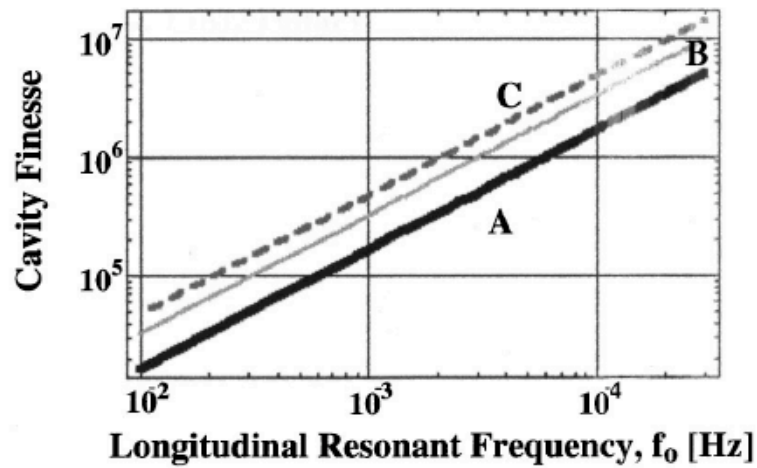

FIG. 11. Required cavity finesse of the noise cancelation cavity as a function of the fundamental longitudinal resonance. A—sapphire with $v=10$ km/s; B—aluminium 5056 with $v=5.1$ km/s; C—niobium with $v=3.4$ km/s.

The resonant frequency of the phase noise cancellation cavity sets an upper limit to the detection of the common-mode free-mass motion induced by gravitational waves. By implementing a sapphire cavity with start-of-the-art mirrors the resonant frequency may be kept as high as $10^4$ Hz. Ground-based kilometre interferometers such as VIRGO and LIGO expect to detect sources in the $10\text{-}10^4$ Hz frequency band, and thus the technique is practical with current technology. Key sources in this frequency range are black hole births, vibrating and precessing neutron stars, stellar core collapse and coalescences of binaries. If any of the above processes are largely symmetrical it may be possible to detect a scalar component of the radiation if it exists.

## VII. CONCLUSION

We have shown how the different types of possible gravitational waves will excite a free-mass system in the local frame. In particular, the knowledge of both the differential and common mode induced signals will provide important information to determine the direction of the incoming signal and help determine between different theories of gravitation. Also, It was shown that a differential detector such as a Michelson interferometer can detect the possible types of gravitational radiation depending on the direction.

To detect the common-mode motion of the free-mass system, the conventional differential interferometric gravitational wave detector can be reconfigured. This is done by simply rotating the beam splitter by $90^0$ and is an adaptation of the Fox-Smith Interferometer. In addition, there is no need to create separate F-P resonance in each arm to obtain multiple-pass enhancement in the sensitivity. All that is needed is a high reflectivity beam splitter to produce a high finesse F-S cavity. The disadvantage to this scheme is the sensitivity to phase noise. A methods to cancel the phase noise without cancelling the signal was proposed. Given that the LIGO project will operate with multi beams in the detector, this configuration could be set up in parallel to the conventional differential detection scheme.

## ACKNOWLEDGMENTS

One of the authors, M.E. Tobar, would like to thank P.R. Saulson for helping him to understand some aspects of laser interferometers via email. The authors would like to thank the members of the TAMA collaboration for many useful discussions during the 1998 February workshop in Kamioka. Also, the authors thank Prof. T. Shintomi and Prof. A. Yamamoto for providing the facilities at KEK High Energy Accelerator Research Organization. This work was funded by the Monbusho Grant-in-Aid for Fellows supported by the Japan Society for the Promotion of Science.

## REFERENCES

[1] A. Abramovici et al, Science **256**, 325 (1992).

2  C. Bradaschia, et al, Nucl. Instrum. and methods A **289**, 518 (1990).

3  http://phwave.phys.lsu.edu/www/users/others/others.html .

4  K. Tsubono, M. K. Fujimoto, and K. Kuroda, (Universal Academy Press Inc., Tokyo, 1997).

5  T. Uchiyama, D. Tatsumi, T. Tomaru, *et al*., Phys. Let. A **242**, 211-214 (1998).

6  P. R. Saulson, *Fundamentals of Interferometric Gravitational Wave Detectors* (World Scientific, 1994).

7  P. J. Steinhardt and F. S. Accetta, Phys. Rev. Lett. **64**, 2740 (1990).

8  C. G. Callan, D. Friedan, E. J. Martinec, *et al*., Nucl. Phys. B **262**, 593 (1985).

9  T. Damour and A. M. Polyakov, Nucl. Phys. B **423**, 532 (1994).

10  T. Damour and A. M. Polyakov, Gen. Relativ. Gravit. **26**, 1171 (1994).

11  C. Brans and R. H. Dicke, Phys. Rev. **124**, 925 (1961).

12  T. Matsuda and H. Nariai, Prog. Theor. Phys. **49**, 1195 (1972).

13  M. Shibata, K. Nakao, and T. Nakamura, Phys. Rev. D **50**, 7304 (1994).

14  M. A. Scheel, S. L. Shapiro, and S. A. Teukolsky, Phys. Rev. D **51**, 4236 (1995).

15  M. Saijo, H. Shinkai, and K. Maeda, Phys. Rev. D **56**, 785 (1997).

16  T. Harada, T. Chiba, K. Nakao, *et al*., Phys. Rev. D **55**, 2024 (1997).

17  R. L. Forward, General Relativity and Gravitation **2**, 149-159 (1971).

18  W. W. Johnson and S. M. Merkowitz., Phys. Rev. Let. **70**, 2367 (1993).

19  M. Bianchi, E. Coccia, C. N. Colacino, *et al*., Class. Quantum Grav. **13**, 2865-2873 (1996).

20  R. L. Forward, Phys. Rev. D. **17**, 379-390 (1978).

21  B. F. Schutz and M. Tinto, Monthly Notices of the Royal Astronomical Society **224**, 131 (1987).

22  C. M. Will, *Theory and Experiment in Gravitational Physics* (Cambridge University Press, Cambridge, 1993).

23  H. Goldstein, *Classical Mechanics* (Addison-Wesley, Cambridge Mass., 1956).

24  Y. Gürsel and M. Tinto, Phys. Rev. D **40**, 3884 (1989).

25  D. G. Blair, E. N. Ivanov, M. E. Tobar, *et al*., Physics Review Letters **74**, 1908-1911 (1995).

26 H. Peng, D. G. Blair, and E. N. Ivanov, Journal of Physics D: Applied Physics **27**, 1150 (1994).

27 LIGO Project, National Science Foundation Technical Review (California Institute of Technology, Massachusetts Institute of Technology, 1995).

28 G. Rempe, R. J. Thompson, H. J. Kimble, *et al.*, Optics Letters **17**, 363-365 (1992).

29 F. Ricci and A. Brillet, Annu. Rev. Nucl. Part. Sci. **47**, 111-156 (1997).

30 D. G. Blair, L. Ju, and M. Notcutt, Rev. Sci. Instrum. **64**, 1899 (1993).

31 L. Ju, M. Notcutt, D. G. Blair, *et al.*, Phys. Let. A **218**, 197 (1996).

32 K. S. Bagdasarov, V. B. Braginskii, and V. P. Mitrofanov, Sov. Phys. Crystallogr. **19**, 549 (1975).

33 R. Storz, C. Braxmaier, K. Jäck, *et al.*, Optics Let. **23**, 1031-1033 (1998).

34 C. R. Locke, M. E. Tobar, E. N. Ivanov, *et al.*, submitted to J. Appl. Phys. (1998).